\documentclass[useAMS,usenatbib]{mn2e}
\usepackage{graphicx}
\usepackage{natbib}
\def\simge{\mathrel{%
   \rlap{\raise 0.511ex \hbox{$>$}}{\lower 0.511ex \hbox{$\sim$}}}}
\def\simle{\mathrel{
   \rlap{\raise 0.511ex \hbox{$<$}}{\lower 0.511ex \hbox{$\sim$}}}}

\usepackage{amsmath}
\usepackage{amssymb}
\usepackage{graphicx}

\title[Cores]{The dynamics of collapsing cores and star formation}
\author[Keto, Caselli, Rawlings]{Eric Keto$^{1}$\thanks{E-mail:
keto@cfa.harvard.edu (EK); \hfill\break
p.caselli@leeds.ac.uk (PC)}, Paola
Caselli$^{2}$, Jonathan Rawlings$^{3}$
\\
$^{1}$Harvard-Smithsonian Center for Astrophysics, 160 Garden St, Cambridge, MA 02420, USA \\
$^{2}$Max Planck Institute for Extraterrestrial Physics, Glessenbachstr. 1, D-85741, Garching, Germany \\
$^{3}$University College London, Gower St, London WC1E 6BT, United Kingdom
}
\begin{document}

\date{October 21, 2014}

%\pagerange{\pageref{firstpage}--\pageref{lastpage}} \pubyear{2009}

\maketitle

\label{firstpage}

\begin{abstract}
Low-mass stars are generally understood to form by the gravitational collapse of 
the dense molecular clouds known as starless cores.  Continuum observations 
have not been able to distinguish among the several different hypotheses that describe
the collapse because the
predicted density distributions are the almost the same, as they are
for all spherical self-gravitating clouds. However, the predicted
contraction velocities are different enough that the models can be discriminated
by comparing the velocities at large and small radii. This can be
done by observing at least two different molecular line transitions that are
excited at different densities. For example, the spectral lines of 
the H$_2$O $(1_{10} - 1_{01})$ and C$^{18}$O $(1-0)$ have 
critical densities for collisional de-excitation that differ by 5 orders of
magnitude. We compare observations of these lines from the contracting
starless core L1544 against the spectra predicted for several different hypothetical models
of contraction including the Larson-Penston flow, the inside-out collapse of the
singular isothermal sphere, 
the quasi-equilibrium contraction of an unstable Bonnor-Ebert sphere, and the
non-equilibrium collapse of an over-dense Bonnor-Ebert sphere. Only the model of the
unstable quasi-equilibrium Bonnor-Ebert  sphere is able to produce the observed shapes of both
spectral lines. This model allows us to interpret other observations of molecular lines in
L1544 to find that the inward velocities seen in observations
of CS$(2-1)$ and N$_2$H$^+$ are located within the starless core itself, in particular
in the region where the density profile follows an inverse square law. If these
conclusions were to hold in the analysis of other starless cores, this would
imply that the formation of hydrostatic clouds within the turbulent
interstellar medium is not only possible but not exceptional
and may be an evolutionary phase in low-mass star formation.

\end{abstract}

\section{Introduction}

The hypothesis that starless cores are the future birthplaces of low mass stars
\citep{Myers1983,MyersBenson1983,Beichman1986} has  proven 
an enduring point amidst continuing revision
in the theories of how 
these cores actually evolve toward star formation. For example,
one of the earlier suggestions that the gravitational collapse of star-forming clouds
such as the starless cores 
occurs via a  
Larson-Penston (LP) flow \citep{Larson1969,Penston1969}, was later
criticized \citep[e.g.][]{Shu1977} because of its
very particular initial conditions. In the LP solution, an initial velocity of 3.3 times 
the sound speed is required to pass a critical point in the flow.
\citet{Shu1977} proposed an
alternative initial state, the singular isothermal sphere (SIS), 
leading to a different evolution described as "inside-out" collapse. This hypothesis
was in turn criticized on similar grounds, namely that the (SIS) is unrealizable as
an initial state owing to its dynamical instability \citep{Whitworth1996}. Alternatively
it was proposed that
the cores may be magnetically rather than hydrodynamically,
supported against their self-gravity. Contraction then proceeds by ambipolar
diffusion \citep{Shu1987,Mouschovias2001,Li1999}. 
This hypothesis faces challenges
in that observations indicate that cores are supported 
predominantly by thermal pressure \citep{Nakano1998}
and furthermore the magnetic diffusion 
time scale may be too slow \citep{Tafalla1998,Ward-Thompson1999}.
Another hypothesis suggests that the cores are 
hydrostatic or Bonnor-Ebert spheres (BES) partially supported by 
turbulent kinetic energy.   Damping of this turbulent kinetic energy \citep{Nakano1998}, 
which might be in equipartition with magnetic wave energy \citep{MyersGoodman1988},
allows the cores to contract in quasi-equilibrium (QE) 
\citep{KC10, Broderick2010}.
The model of the cores as BES
has been criticized as incompatible with the larger-scale supersonic
turbulence of the interstellar medium (ISM)  \citep{BP2003}.  
In this view, the turbulence creates 
transitory zones of compression where the distribution of 
dense gas merely resembles 
that in hydrostatic cores
\citep{MacLowKlessen2004}.

The scientific debate continues. \citet{FosterChev1993} have
suggested that an LP flow can develop in the contraction of a 
sphere whose initial state is both out of equilibrium and gravitationally unstable.
 \citet{CiolekBasu2000} suggest how to 
reconcile the hypothesis of ambipolar diffusion with observations. The 
BES description, enhanced to include non-isothermal gas \citep{Evans2001}
chemical abundance variations \citep{KC08},
and perturbative oscillations (sonic waves) \citep{Keto2006, Broderick2010},
continues to prove both descriptive and predictive of a wide range of observations 
\citep{Alves2001,Tafalla2002,Lada2003,Caselli2010,Caselli2012,Nielbock2012}.

\begin{figure} 
$
\begin{array}{cc}
\includegraphics[width=3.25in]{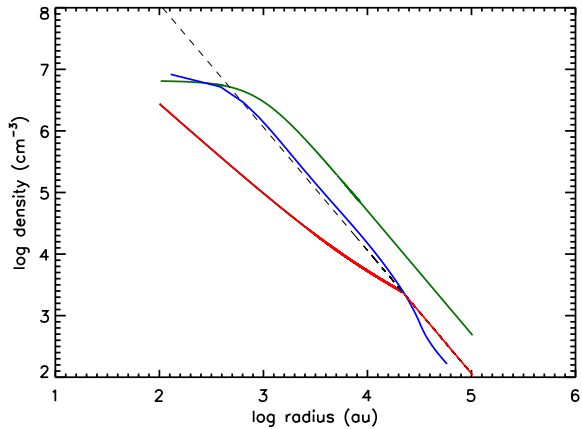}
\end{array}
$
\caption{
Densities ({\it right}) in 3 models of spherical contraction. The green line (upper) shows
the Larson-Penston flow. The red line (lower) shows the inside-out collapse of a singular isothermal sphere.
The blue line (middle) shows the quasi-equilibrium contraction of an unstable Bonnor-Ebert sphere. The dashed line
 shows a $r^{-2}$  profile for reference.
}
\label{fig:vd1a}
\end{figure}

One difficulty in discriminating among these hypotheses for 
contraction toward star formation is that some of
the observed properties of the cores can be explained by more than one 
theoretical model. For example, self-gravitating
spherical clouds develop a density profile scaling as the inverse square of
the radius under a wide range of conditions. 
Figure \ref{fig:vd1a} shows the gas densities in the three
different evolutionary models with spherical symmetry,
the LP flow, the inside-out collapse of the SIS, and the
contraction of a QE-BES. The different density profiles are close enough to a $1/r^2$
profile to be observationally indistinguishable.
The non-spherical models, ambipolar diffusion
and large-scale turbulence can also produce
similar density profiles \citep{BP2003, CiolekBasu2000} but a detailed
comparison with these more complex models is beyond the scope of this paper.

\begin{figure} 
$
\begin{array}{cc}
\includegraphics[width=3.25in]{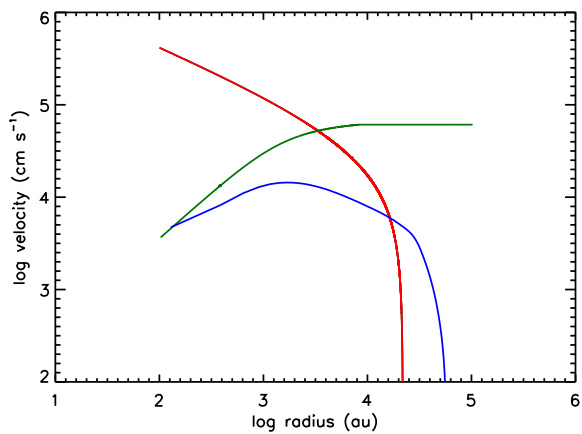}
\end{array}
$
\caption{
Velocities in the same 3 models of spherical contraction as in figure \ref{fig:vd1a}. In contrast
to the density profiles, the velocity profiles
of these 3 models are sufficiently different to  be observationally distinguishable. The green line (middle) shows
the Larson-Penston flow, the red line (upper) shows the inside-out collapse of a singular isothermal sphere, and
the blue line (lower) shows the quasi-equilibrium contraction of an unstable Bonnor-Ebert sphere. }
\label{fig:vd1b}
\end{figure}

Although the density profiles are similar in all self-gravitating spherical clouds
regardless of the mode of contraction, the internal velocities are quite different. 
Figure \ref{fig:vd1b} shows that the velocities in the
LP, SIS, and QE-BES models are sufficiently
different that the underlying model can be deduced from the
observations simply by comparing velocities at large and small radii.
While the velocities in the spherical hydrodynamic models are quite specific,
it is difficult to specify velocities for the hypotheses of turbulent compression 
or ambipolar  diffusion. In the former case, the velocity profiles 
can be quite varied \citep{BP2003} because of
the random nature of turbulence. In the latter case, the number of free or
unconstrained parameters including the geometry of the magnetic field and the spatial
distribution of magnetic field strengths allows for diverse outcomes \citep{Li1999,CiolekBasu2000}.
In this paper we will set aside these last two hypotheses to await more statistical tests and focus
on discriminating among the velocity profiles of the more uniquely defined
spherical models.

Because of the approximate $1/r^2$ relationship between density and 
radius in all three spherical models, it is possible to measure the contraction 
velocity at small and large radii by observing spectral
lines from molecules that are excited or abundant at different densities. 
We have observations of two such molecular lines for the contracting core L1544.
Owing to its large Einstein A coefficient ($3.5 \times 10^{-3}$ s$^{-1}$), the 556.936 GHz 
emission line of H$_2$O ($1_{10}-1_{01}$) serves as a tracer of dense gas 
($\simge 10^6$ cm$^{-3}$) that isolates the conditions in the very center ($r < 0.01$ pc) of the 
core \citep{KRC14}. In contrast,  the 109.782 GHz C$^{18}$O (1-0) transition has a critical density
that is 5 orders of magnitude smaller. Thus the observed C$^{18}$O line is dominated by the
emission from the outer regions of the cloud where there is more volume
and more column density. 

In this paper, we compare observations of these two spectral lines from the contracting core L1544
against the line spectra predicted for  5 different spherical models as follows:
\begin{enumerate}
\item The contraction of a quasi-equilibrium Bonnor-Ebert sphere (QE-BES) \citep{KC10}.
\item The contraction of a non-equilibrium Bonnor-Ebert sphere (NE-BES) \citep{FosterChev1993}.
\item A static sphere.
\item The Larson-Penston (LP) flow \citep{Larson1969,Penston1969}.
\item  The inside-out collapse of the singular isothermal sphere (SIS) \citep{Shu1977}.
\end{enumerate}

Since a BES
is in hydrostatic equilibrium by definition, the terminology of a quasi-equilibrium BES or 
non-equilibrium BES sounds self-contradictory and requires some explanation. 
Both the QE-BES and NE-BES are approximate BES related to a true BES
by construction. The QE-BES is constructed as an unstable BES and allowed to 
evolve hydrodynamically. In the early stages, the QE-BES resembles a BES except
in the center which evolves out of equilibrium most rapidly (figure \ref{fig:vd2}). The NE-BES is
constructed from an unstable or nearly unstable BES that is  given a 10\% increase in density
everywhere and allowed to evolve. Because it is out of equilibrium everywhere,
the contraction of NE-BES proceeds differently
from a QE-BES.  It begins collapse on all scales at once (figure \ref{fig:vd2}).

Our investigation finds that the contracting QE-BES model
provides a reasonable match to the observed emission 
lines H$_2$O ($1_{10}-1_{01}$)  and C$^{18}$O (1-0) of the starless core L1544.
The comparison with the other models indicates how close the
contraction in L1544 is to quasi-equilibrium. For example, the QE-BES, the NE-BES,
and the SIS all have an inside-out type of contraction with higher velocities toward
the center, but with observable differences.
For example, the NE-BES produces a C$^{18}$O 
spectral line profile that does not fit the observed C$^{18}$O line because its
overall contraction has
higher velocities in the outer core that split the predicted C$^{18}$O line in a way 
that is not observed. 
The SIS with its continuously increasing
velocities at smaller radii predicts an H$_2$O line 
with a larger split than observed.  

In this paper we also use the conceptual framework of the QE-BES model
to interpret other observations in the literature bearing on
the evolution of the starless cores towards star formation. 
In particular, we model the CS(2-1) and N$_2$H$^+$(1-0) emission
in L1544 to precisely locate the "extended inward motions" \citep{Tafalla1998,Myers2000}
in starless cores. We find that the N$_2$H$^+$ emission comes from the central 5000 au (half-width
radius at half-maximum (HWHM)) where the gas density is $\simge 10^5$ cm$^{-3}$.
The CS(2-1) emission comes from within 10,000 au (HWHM) which is within the $1/r^2$ region of the 
QE-BES density profile. 
In particular, we find that the comparison of the CS and N$_2$H$^+$ lines does not support an
interpretation of inward flow from scales larger than the starless core.
%-----------------------------------------------------------
%FIGURE 1
% /sma/ketoSci2/projects/current/h2o-dynamics/shu/shu77/triple.pro
\break
%-----------------------------------------------------------
%-----------------------------------------------------------
%FIGURE 2
% /sma/ketoSci2/projects/current/h2o-dynamics/shu/shu77/triple.pro
\break
%-----------------------------------------------------------

\section{Data}

The H$_2$O and CO spectra are the same data as used in \citet{KC10} and \citet{KRC14}. 
The water line was 
observed with the Herschel Space Observatory and 
originally reported in \citet{Caselli2012}. 
The C$^{18}$O line was observed with the IRAM 30m and 
originally reported in \citet{Caselli1999}.

Observations of the CS(2-1) line are reported 
in the literature \citep{Tafalla1998,Myers2000,Lee2001} and
provide further information on the nature of the contraction. 
This transition has a critical density of $6\times 10^5$ cm$^{-3}$ 
intermediate between H$_2$O and C$^{18}$O. 
Unlike the C$^{18}$O, the CS transition is 
optically thick in L1544. Differential self-absorption of the line
emission in a velocity gradient makes this transition a sensitive
indicator of velocity.  We refer to the CS(2-1) spectra in the
literature in a qualitative comparison with the predicted spectra.

\section{The dynamical models}

Our BES models which 
include the QE-BES, NE-BES and the 
static BES are calculated with the 1D
Lagrangian numerical hydrodynamic code described 
in \citet{KF05}. This code includes radiative heating
and cooling of the gas and dust and simplified CO and 
H$_2$O chemistry to set the abundances
of the major gas coolants \citep{KRC14}. 
These abundances are also used 
in the radiative transfer calculations that
simulate the observed spectral lines.

The QE-BES model starts with a central
density of $10^4$ cm$^{-3}$, and the Lane-Emden 
density profile is truncated -- the
external pressure is set equal to the internal pressure 
at the boundary -- at a radius 
that gives the cloud a mass of 10 M$_\odot$. This is
the same model used in our earlier research \citep{KC08,KC10}.
The modeled mass is approximately the same as the 8 M$_\odot$
estimated from CO
observations of L1544 itself, excluding the L1544-E and L1544-W
companions \citep{Tafalla1998}. The initial density at the start of
the hydrodynamic evolution cannot be known from observations.
Ours is chosen to be high enough that the core is gravitationally
unstable and low enough that the core has time to evolve
before reaching the observed central density of about $10^7$ cm$^{-3}$.
This density was deduced 
earlier \citep{KC10} by comparison of the
observed and predicted CO and N$_2$H$^+$ spectra.

\begin{figure}
%$
%\begin{array}{cc}
\includegraphics[width=3.25in]{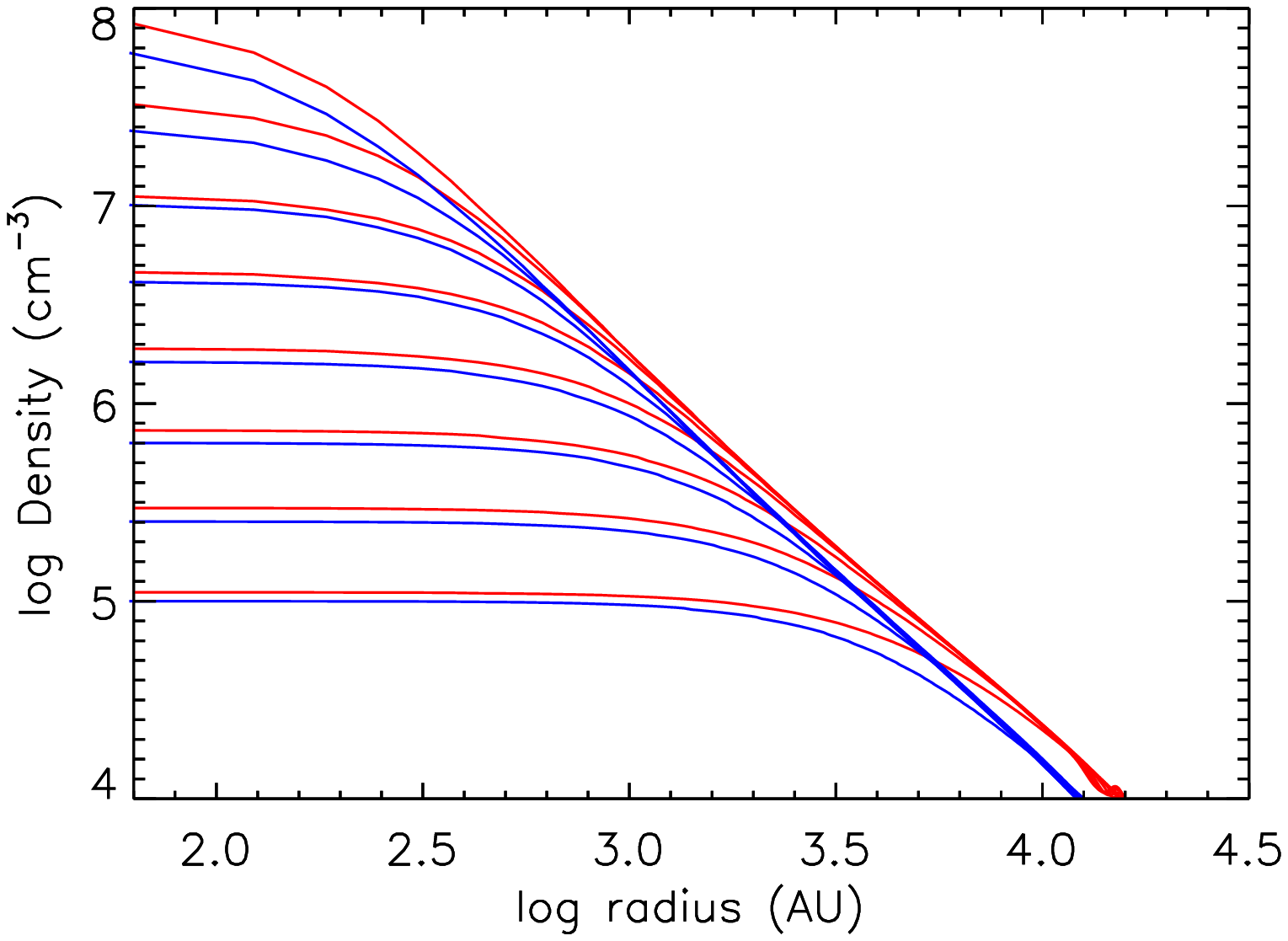}
\includegraphics[width=3.25in]{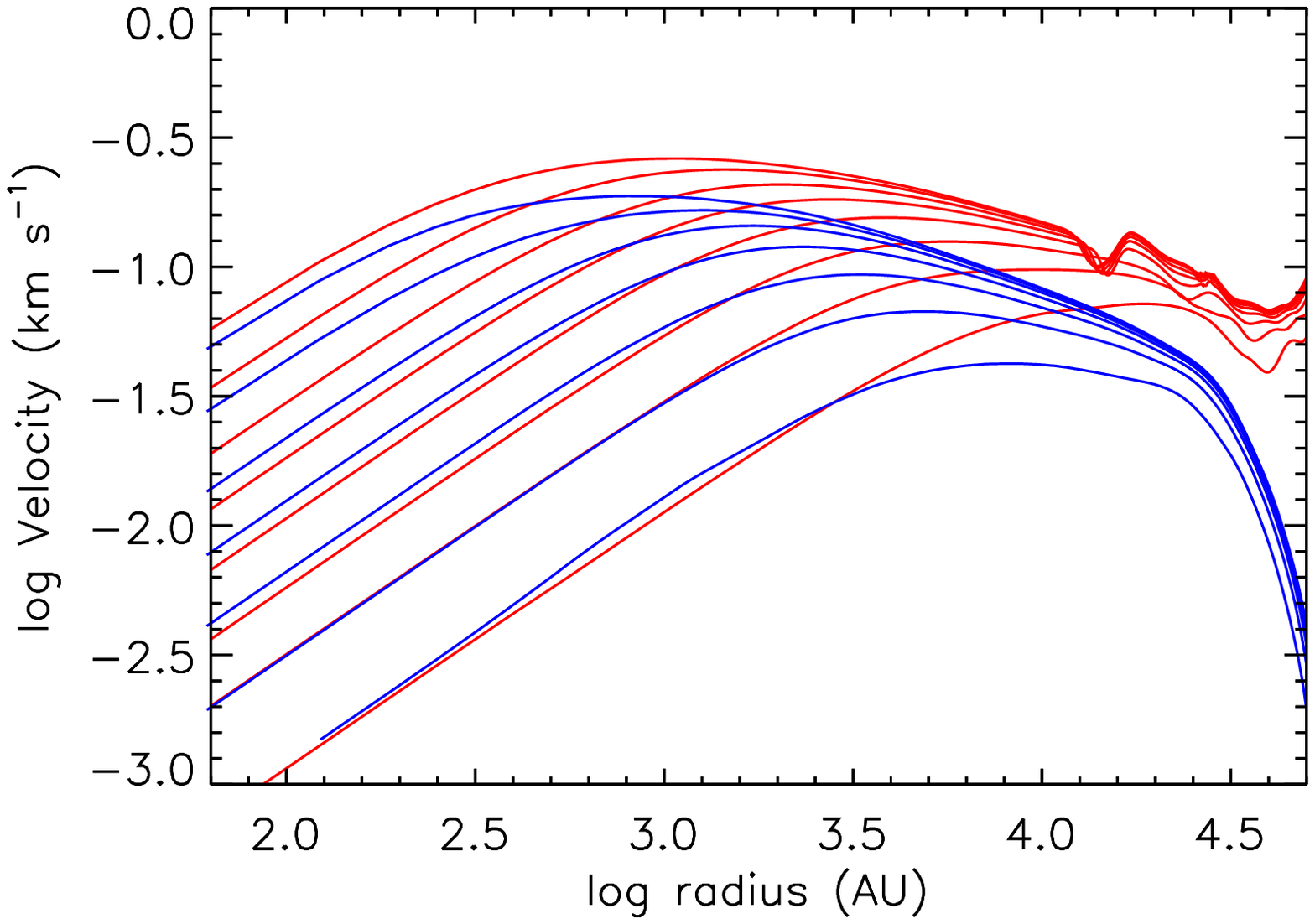}
%\end{array}
%$
\caption{
Comparison of densities ({\it top}) and velocities ({\it bottom}) 
in quasi-equilibrium (QE) ({\it blue}) and non-equilibrium (NE) ({\it red})
contraction at different evolutionary times. The calculation
shows increasing densities and velocities, upward in both
figures as the contraction progresses.
The density structure in both modes of 
contraction continues to resemble a static BES whose
central density is increasing with time. 
The velocities are also nearly the same in the center of the 
core resulting in nearly identical H$_2$O spectra.
However, the velocities are 
quite different in the outer core, tending to
zero in the quasi-equilibrium case and remaining nearly 
constant with both time and radius in the non-equilibrium case. 
This difference
is noticeable in the predicted CO spectra labeled QE-BES and NE-BES
in figure \ref{fig:spectra}. 
}
\label{fig:vd2}
\end{figure}

The initial state is one of
gravitationally unstable hydrostatic equilibrium. 
Any perturbation of this structure will
result in collapse. Even a slight {\it decrease} in 
the overall density will cause the core to
contract because  there
is no adjustment that leads to stable equilibrium. 
The collapse progresses from the inside-out with
the center moving out of equilibrium soonest and fastest. 
The progression of velocities and densities
during the evolution is shown in figure \ref{fig:vd2}. 
We call this mode of contraction quasi-equilibrium (QE)
because wherever the contraction velocities are subsonic,
the density structure resembles a static equilibrium core. 
This is characteristic of an unstable BES during the sonic 
phase of contraction
\citep{Larson1969, FosterChev1993, Ogino1999, KC10}.

We construct an NE-BES by finding the equilibrium 
BES for a 9.1 M$_\odot$ core with the same central
density of $10^4$ cm$^{-3}$ as the QE-BES. 
We then increase the density everywhere by 10\% similar to the
procedure in \citet{FosterChev1993} so 
that the core now has 10 M$_\odot$ but with the density
profile appropriate for a 9.1 M$_\odot$ core. 
This initial state is also gravitationally unstable, but nowhere
in equilibrium. As a result, this core begins to contract 
immediately and everywhere at once. 
Figure \ref{fig:vd2} compares the densities 
and velocities  of the the QE-BES and NE-BES models during the evolution. 
The most significant difference is in the velocities at large radii. 
To compare with the observations we use the model at the evolutionary
time when the central density is $10^7$ cm$^{-3}$, 
same as the model of the QE-BES.

Our third model, the static core, is calculated 
as a 10 M$_\odot$ BES with a central density of $10^7$ cm$^{-3}$. 
The density structure of this core
is quite similar to the QE-BES and the NE-BES  with the
same central densities (figure \ref{fig:vd2}).

The LP and SIS models are calculated by integrating equations 11 and 12 in \citet{Shu1977}. 
The sound speed for both models is 0.2 kms$^{-1}$ ($\sim 11$ K), 
and the evolutionary times are $2.8\times 10^{11}$ and $8\times 10^{12}$ seconds 
respectively, chosen so that the density profiles are qualitatively close to that of the 
contracting BES model (figure \ref{fig:vd1a}). 

The predicted spectral lines for all five models are calculated with our 
radiative transfer code MOLLIE \citep{K90,KR10} which also
computes the chemical abundances \citep{KC08,KRC14} for the LP and SIS
models.  

The LP and SIS models are isothermal in contrast to  
our BES models and in contrast to the observations \citep{Crapsi2007} 
that also indicate a range in temperature from 6 - 18 K in L1544. This 
discrepancy is not a concern for this study. First,
this range in temperature has little effect on the dynamics \citep{KC10}. 
Second, in this study we use the line profiles in our model-data 
comparisons, particularly the splitting of the spectral lines by the velocities, 
rather than the absolute brightness of the 
lines. The former is a function of the gas velocities and the latter of gas temperature.
The correlation between the velocity and gas temperature 
(colder in the center)  affects the spectral line profile, but we will see in 
section \S \ref{comparison} that this is not a problem 
for this conclusions of the comparisons.

% FIGURE 3
% Use the asterisk after figure to make the figure span 2 columns
% /sma/ketoSci2/projects/current/H$_2$O-dynamics/caselli5/structure/density.pro velocity.pro
%\begin{figure*}
%\end{figure*}

\section{The model for the molecular abundances}

\begin{figure}
%$
%\begin{array}{cc}
\includegraphics[width=3.25in]{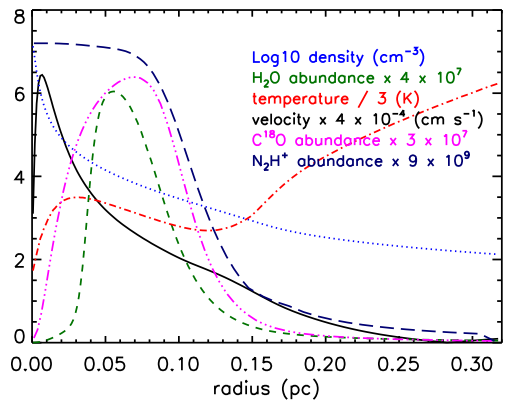}
%\end{array}
%$
\caption{Physical conditions in the QE-BES model for L1544.
The vertical axis has different units depending on the quantity
displayed. 
The solid black line shows the velocity in cms $^{-1}$ multiplied by $4\times 10^{-4}$.
The dotted blue line shows the log of the density (cm$^{-3}$).
The red dot-dash line with 1 dot shows the temperature in K divided by 3.
The green dashed line shows the abundance of H$_2$O relative to H$_2$ multiplied by $4\times 10^7$.
The purple dot-dash line with 3 dots shows the abundance of C$^{18}$O relative to H$_2$ multiplied by $3\times 10^7$. 
The blue line with large dashes is the abundance of N$_2$H$^+$ multiplied by $10^{10}$.
The other models, NE-BES, static BES, LP, and SIS, have similar
temperatures, densities, and abundances, but different velocities.
}
\label{fig:abundances}
\end{figure}

We use simplified models for the gas phase abundances
of CO and H$_2$O following our earlier research 
\citep{KC08,KRC14}. Figure \ref{fig:abundances}
shows the abundances as a function of radius along with the
density, temperature, and velocity of the QE-BES at the evolutionary
time when the central density reaches $10^7$ cm$^{-3}$.  
Both the CO and H$_2$O are depleted from the gas phase 
at high densities 
($\simge 10^5$ cm$^{-3}$) where the molecules are 
frozen onto dust grains. In contrast, the abundance of N$_2$H$^+$
is assumed to remain volatile at all densities across the core.

Despite the gas phase depletion at high density, we know that
that the H$_2$O ($1_{10}-1_{01}$) line emission comes from
the very center of L1544.
The critical density for collisional de-excitation of this transition is
$10^8$ cm$^{-3}$ assuming 10 K and para-H$_2$ \citep{Dubernet2009,KRC14}. 
The maximum gas density, $10^7$ cm$^{-3}$, in 
our model of L1544 never reaches this critical density and
the transition is sub-critically excited \citep{KRC14}.
The volume of gas with high density, 
for example above $10^6$ cm$^{-3}$, is quite small (radii $< 0.005$ pc)
and the H$_2$O abundance is very low  ($ < 10^{-12}$ relative to H$_2$).
The combination of a low column density and sub-critical excitation
means that the H$_2$O ($1_{10}-1_{01}$) emission is very weak (0.025 K),
but serves the purpose of reporting on the conditions in the core
center.

In contrast, the critical density for the C$^{18}$O (1-0) transition ($2\times 10^3$ cm$^{-3}$)
is 5 orders of magnitude smaller than for our H$_2$O transition.
The volume of gas in L1544 with a density high enough to excite
C$^{18}$O is much larger (within 0.14 pc) than the volume able to excite H$_2$O.
Accordingly, the C$^{18}$O (1-0)
line emission is much brighter (5 K) than the H$_2$O line. 
While there is still C$^{18}$O emission from
the center of the core, its
contribution to C$^{18}$O line emission is insignificant.
The C$^{18}$O line is dominated by the large volume and high
column density gas in the outer cloud.

% FIGURE 4
% /sma/ketoSci2/projects/current/h2o-dynamics/caselli5/h2o_std

\section{Comparison and interpretation}\label{comparison}

\subsection{The QE-BES}
\begin{figure*} 
%\begin{figure}
$
\begin{array}{cc}
\includegraphics[width=2.75in]{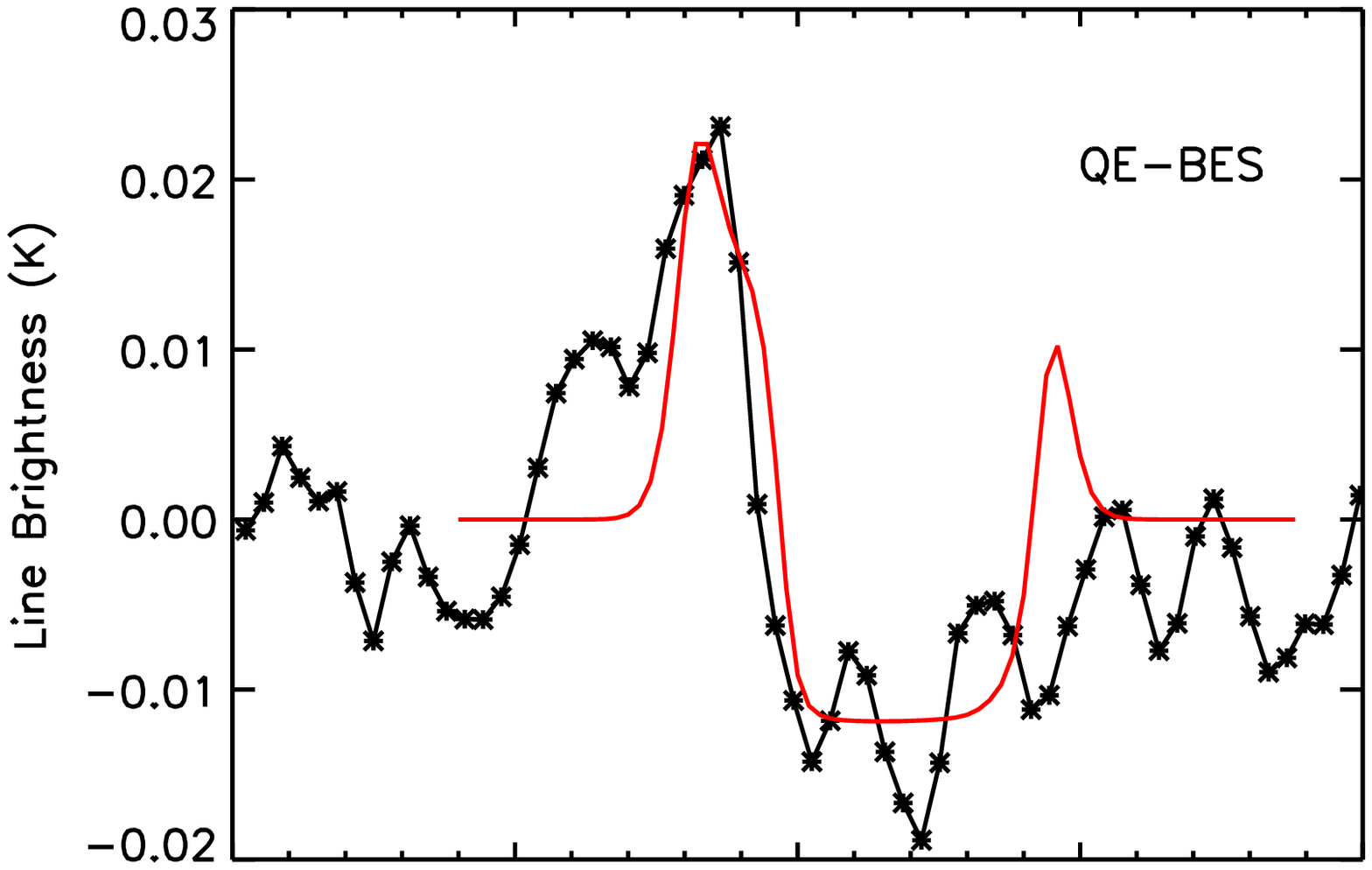}
\includegraphics[width=2.75in]{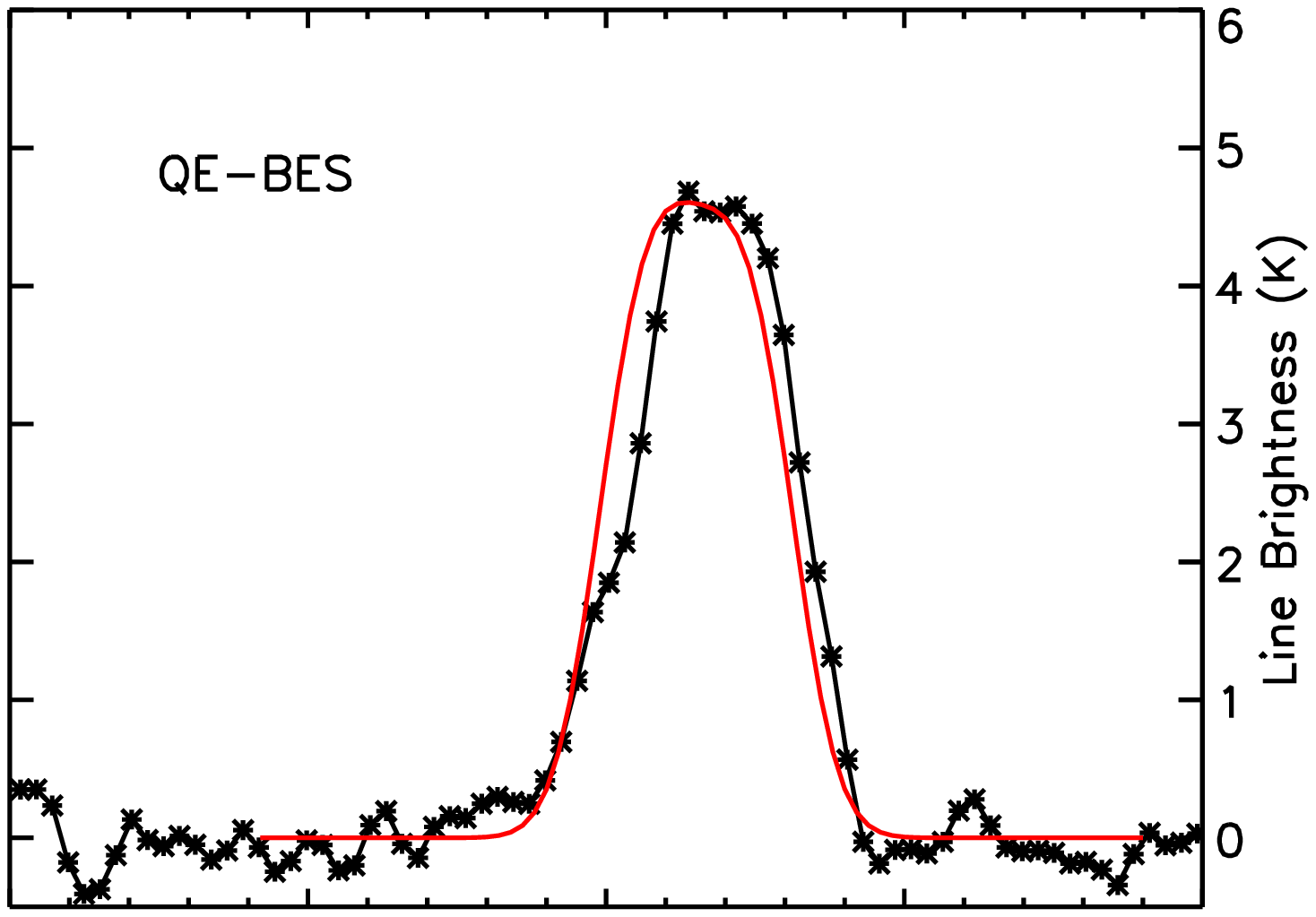} \\
\end{array}
$
\vskip -0.4truein
$
\begin{array}{cc}
\includegraphics[width=2.75in]{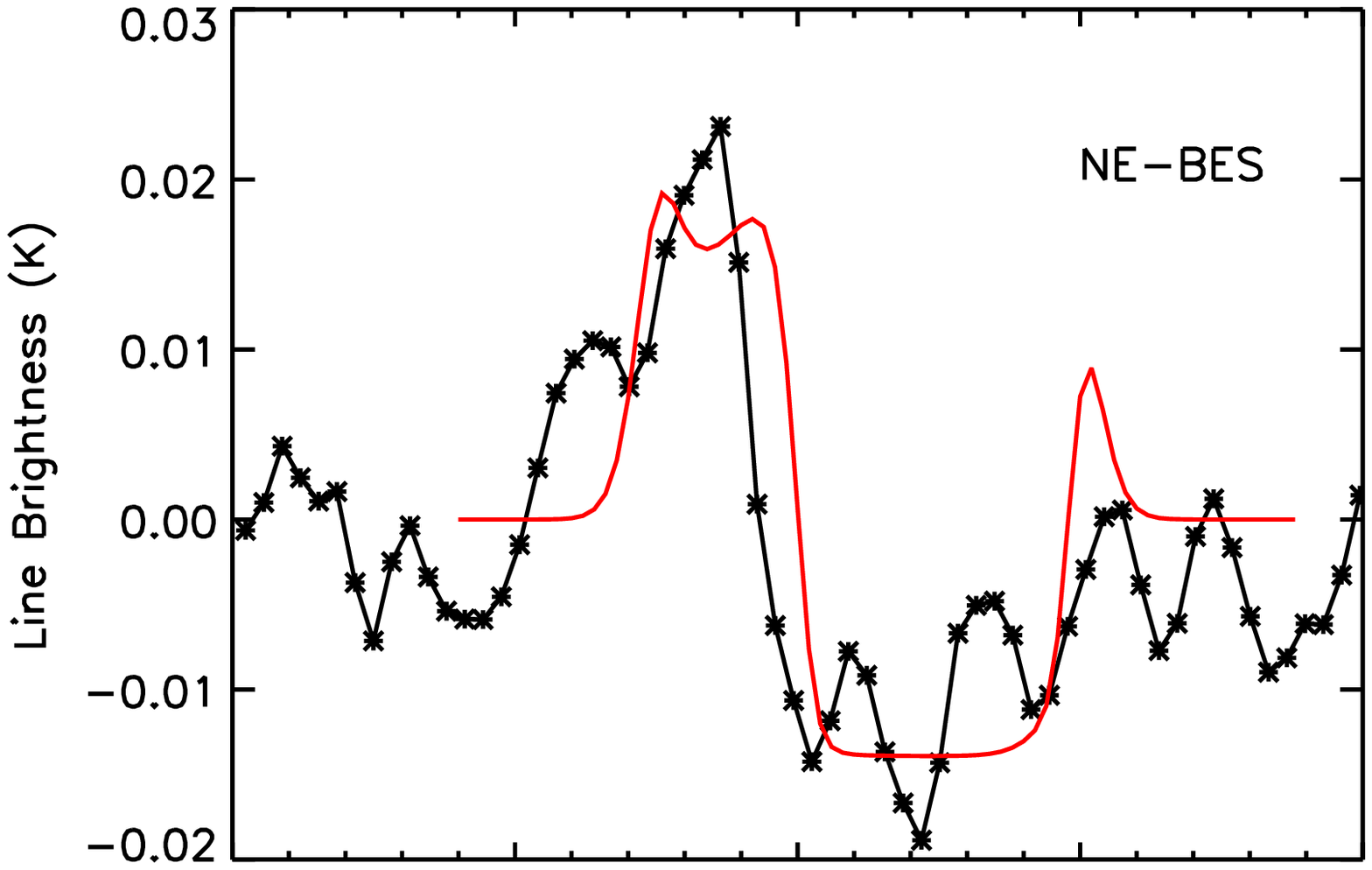}
\includegraphics[width=2.75in]{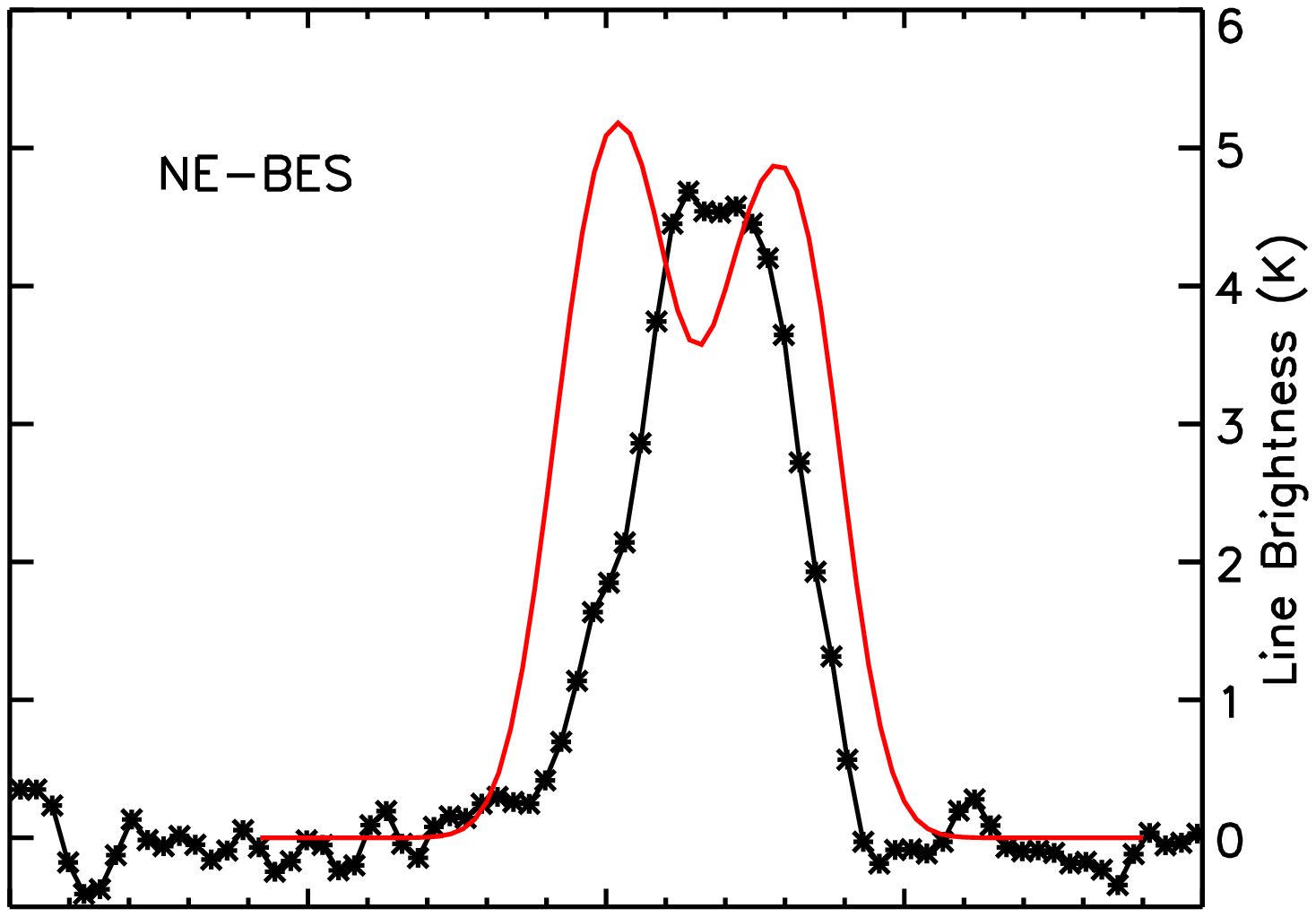} \\
\end{array}
$
\vskip -0.4truein
$
\begin{array}{cc}
\includegraphics[width=2.75in]{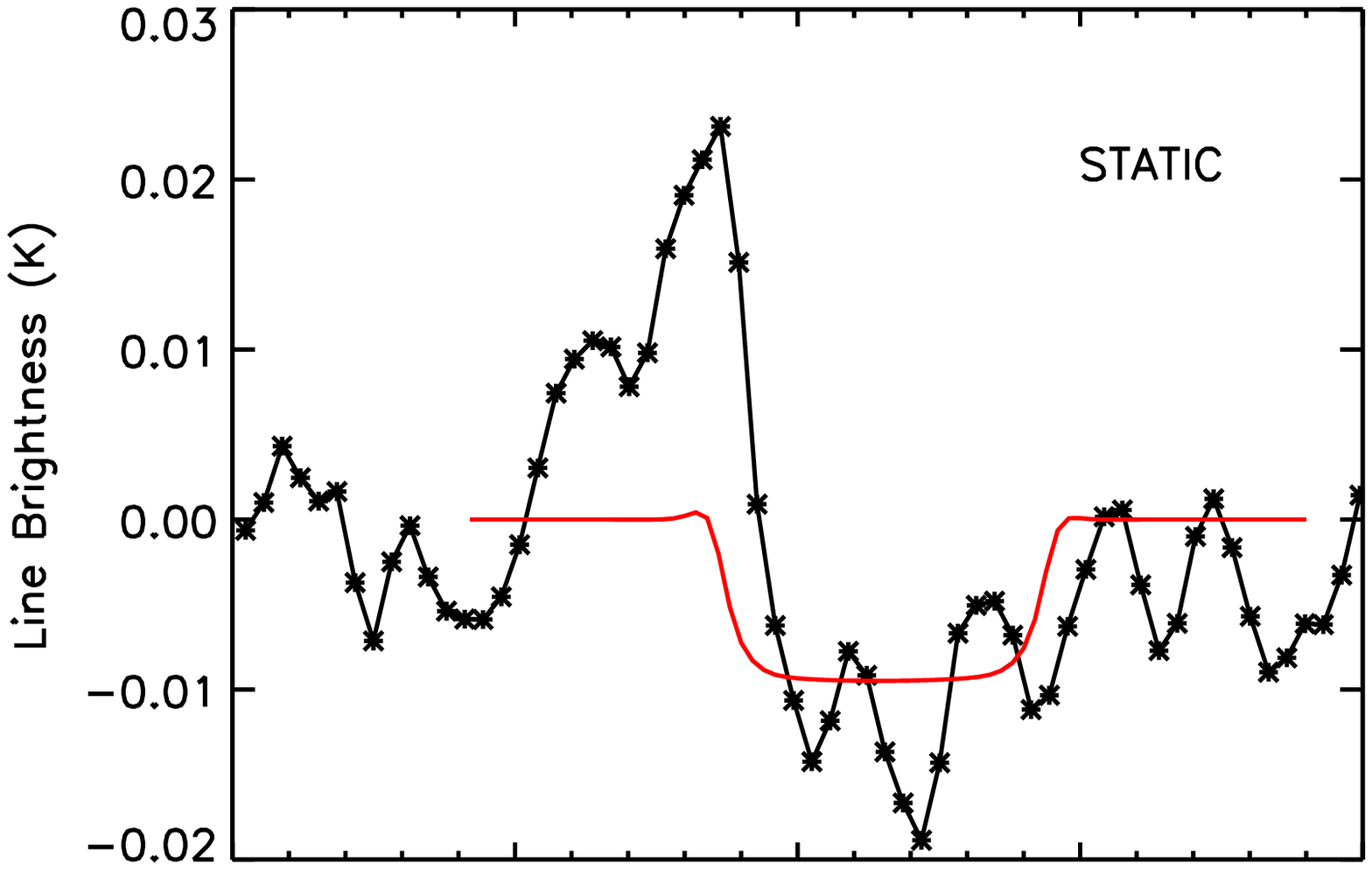}
\includegraphics[width=2.75in]{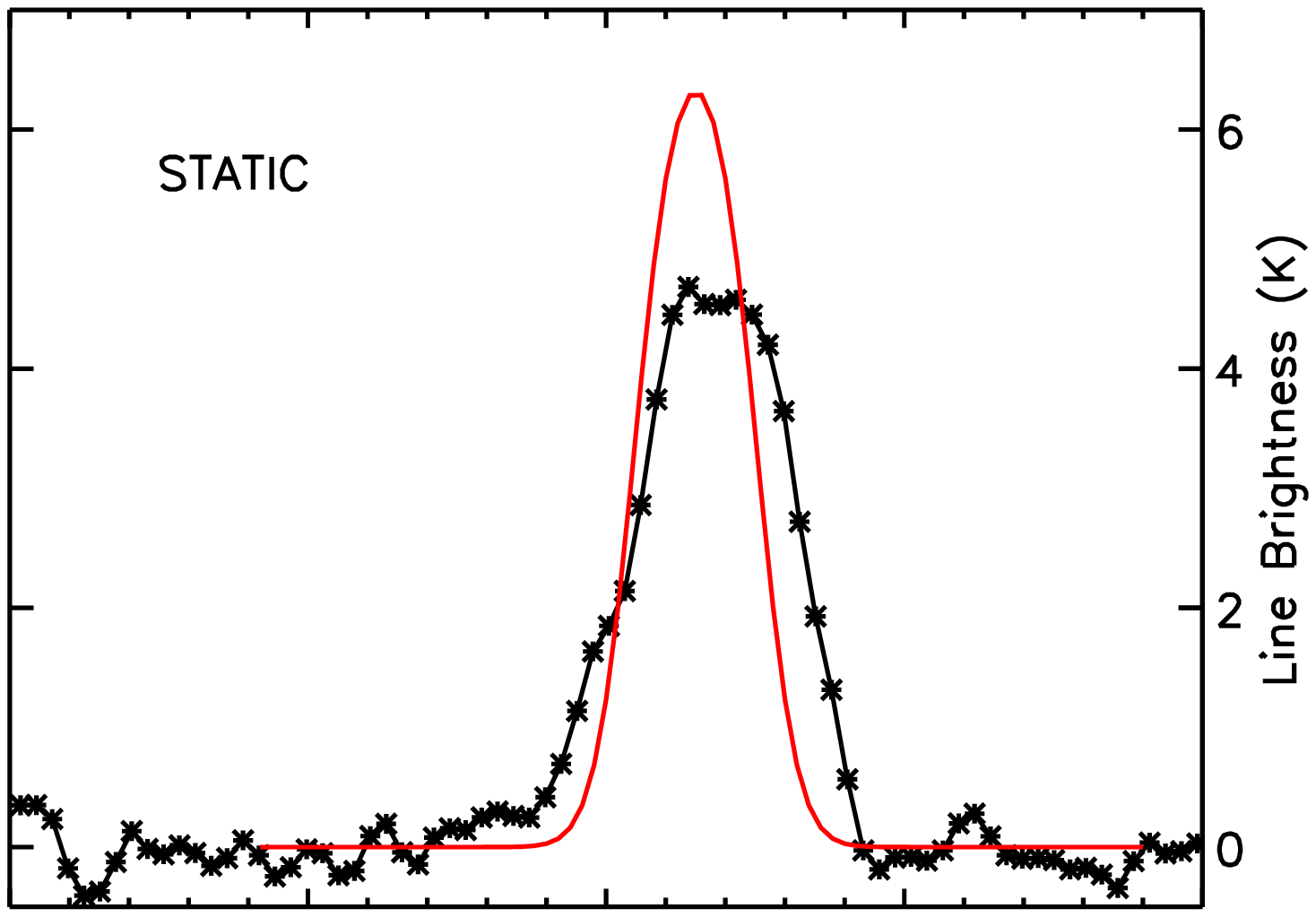} \\
\end{array}
$
\vskip -0.4truein
$
\begin{array}{cc}
\includegraphics[width=2.75in]{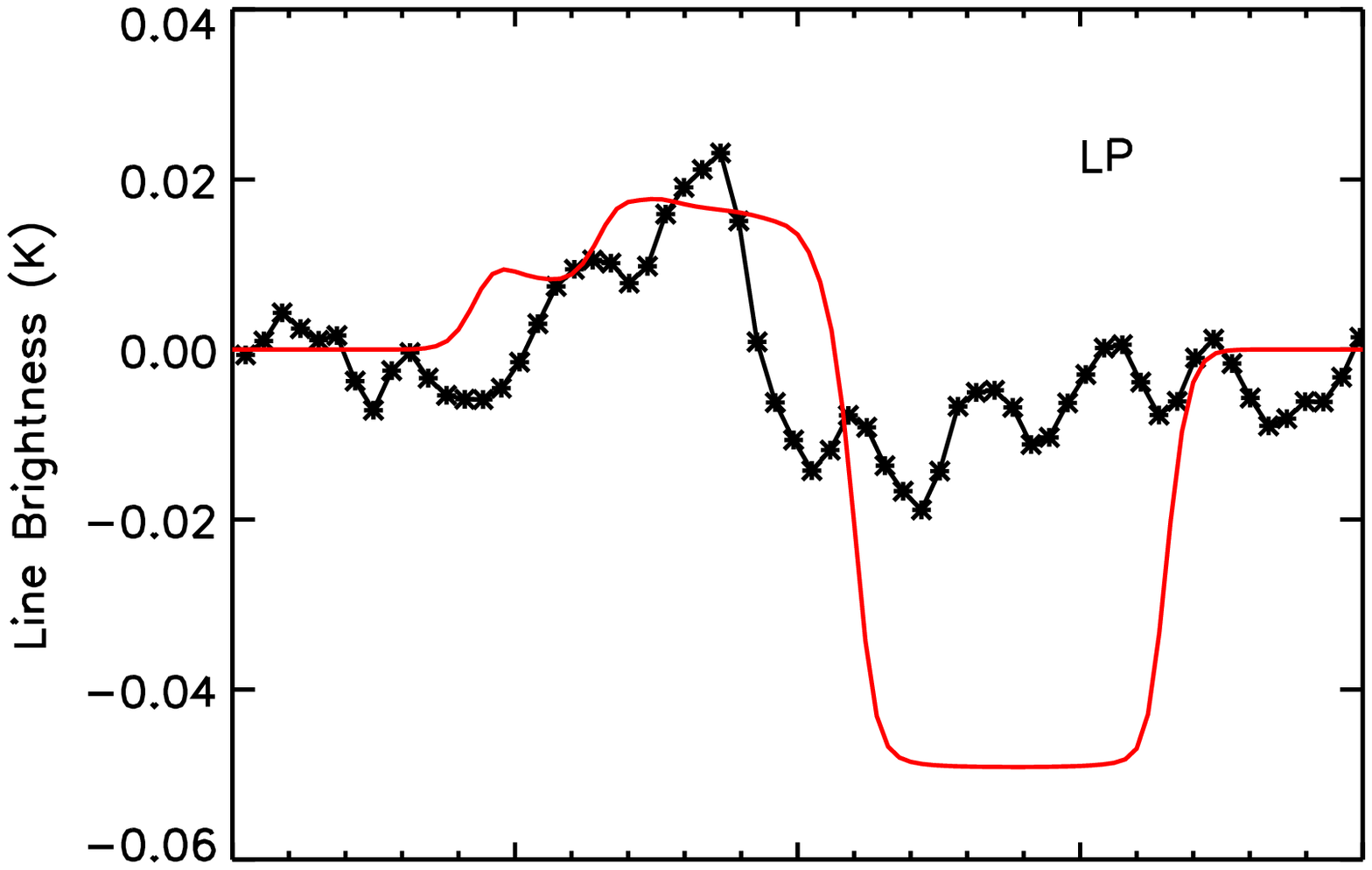}
\includegraphics[width=2.75in]{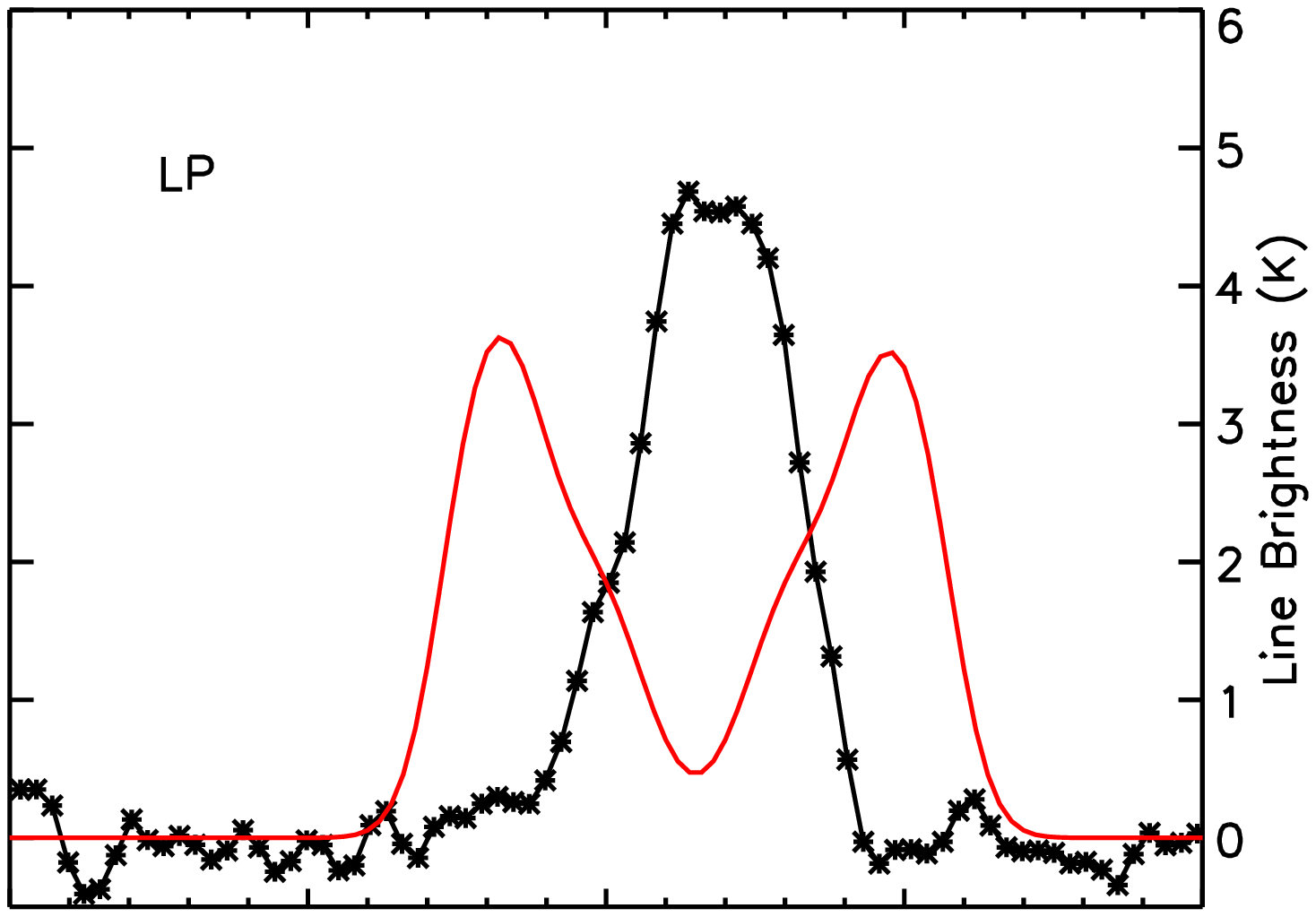} \\
\end{array}
$
\vskip -0.4truein
$
\begin{array}{cc}
\includegraphics[width=2.75in]{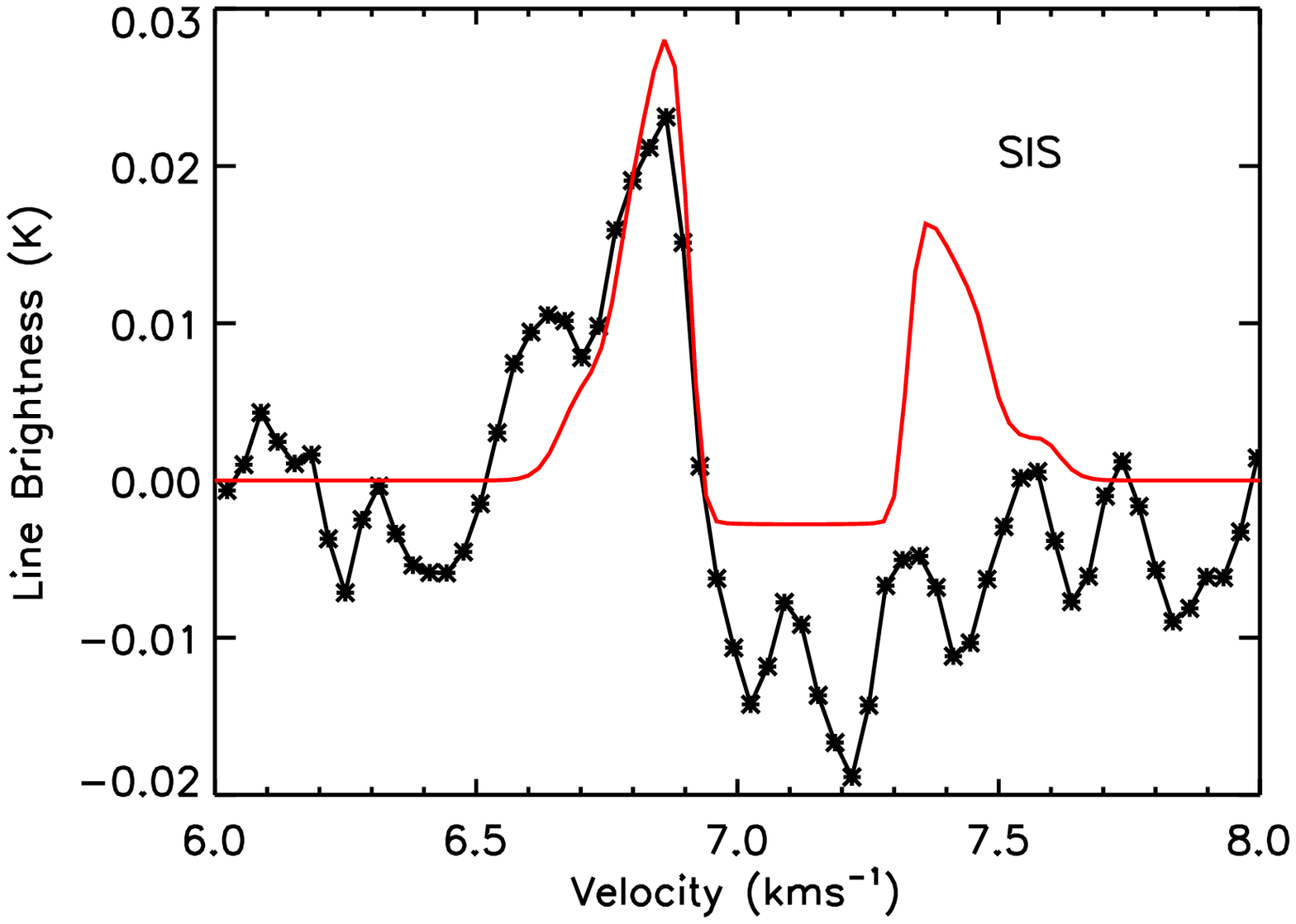}
\includegraphics[width=2.75in]{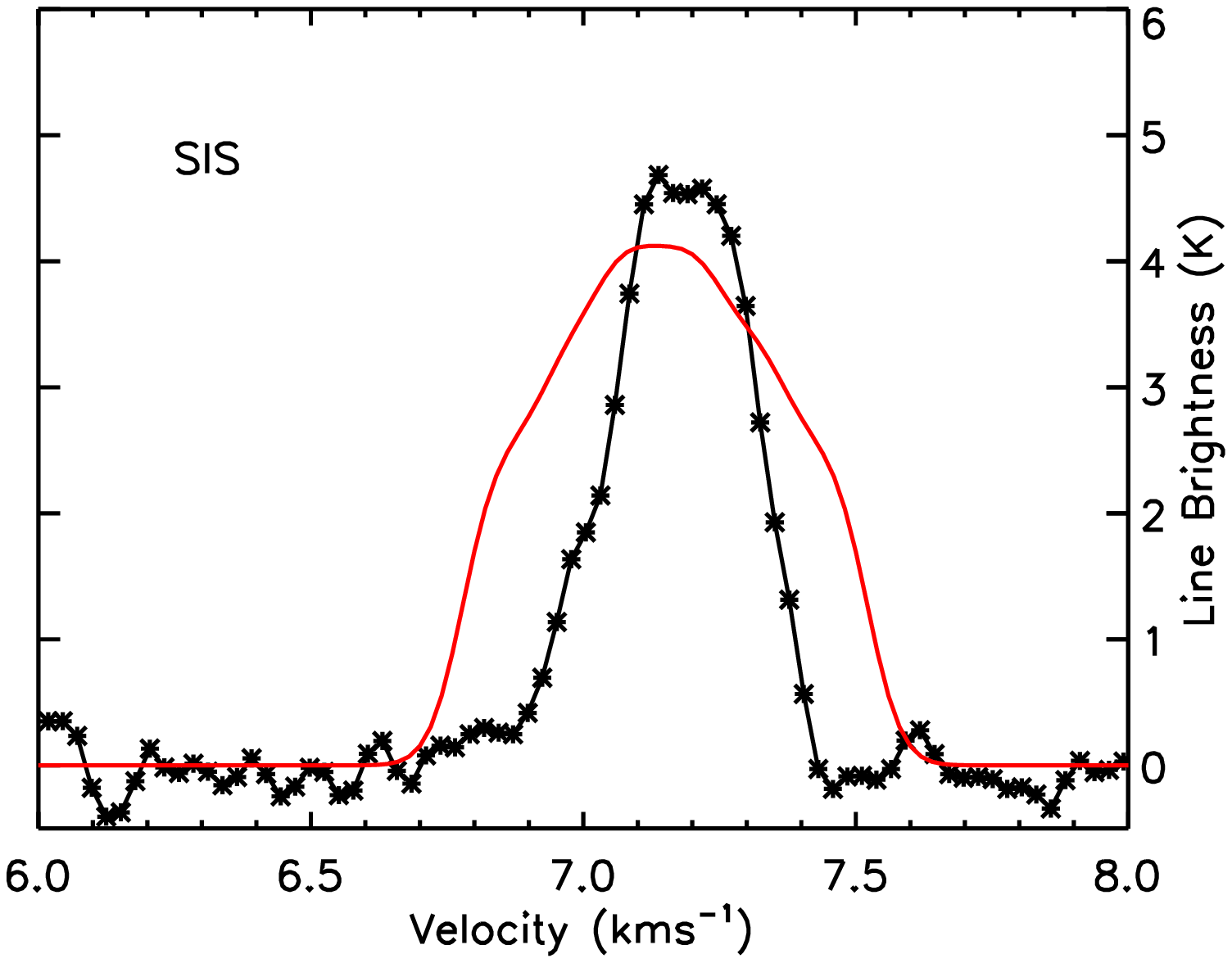}
\end{array}
$
\caption{
Observed spectra (black line with symbols) compared with spectra predicted (plain red line) for the five models of
starless cores. The left and right columns show the H$_2$O 
and C$^{18}$O spectra, respectively.  The five models from top to bottom are:
1) an unstable Bonnor-Ebert sphere in quasi-equlibrium contraction (QE-BES),
2) a Bonnor-Ebert sphere with a 10\% over density in non-equilibrium collapse (NE-BES),
3) a static Bonnor-Ebert sphere (static),
4) a Larson-Penston flow (LP), and 
5) a singular isothermal sphere (SIS) in inside-out collapse. 
Owing to their different critical densities for collisional de-excitation, 
the H$_2$O and C$^{18}$O
lines are generated in the inner and outer regions of the core, respectively.
Only the QE-BES model has the correct contraction
velocities in both regions to produce 
spectral line profiles matching both observations. 
}
\label{fig:spectra}
\end{figure*}

The predicted spectra for the contraction of the QE-BES are 
shown in figure \ref{fig:spectra} and compared with 
the observed spectra. The H$_2$O spectrum shows 
an inverse P-Cygni type profile with red and blue
shifted absorption and emission, respectively, 
consistent with the inward velocities of contraction.
Since this emission must come from the highest density
gas, these velocities are
in the core center. 
In contrast, the CO line, whose emission is dominated by the greater volume of
gas at larger radii, is completely unaffected by the high velocities in the center 
of the core that split the H$_2$O line. 
As a result, the CO line
is only slightly broadened, and this is caused entirely
by the low velocities in the outer part of the QE-BES.

In the H$_2$O spectrum, the absorbed background
is the submillimeter dust continuum, and this has been 
subtracted from both the predicted and observed spectra. 
At the frequency of the CO line the dust continuum 
is negligible so there is no absorption.  

The evolutionary time of the contraction is of course not known,
and we choose a time in the modeled contraction to match the observations.
In our earlier paper \citep{KC10}
we chose the point in the evolution when the 
central density reaches $10^7$ cm$^{-3}$ based on comparison with
CO and N$_2$H$^+$ spectra. 
The observed H$_2$O ($1_{10}-1_{01}$) line provides additional constraint.
Models with much lower central density will not excite the H$_2$O transition
enough to match the observation. 
The observed line profiles and widths 
constrain the model to evolutionary stages before the development
of supersonic velocities in the emitting regions. The maximum
velocity (0.15 kms$^{-1}$) in our selected model is just below the sound speed
(0.2 at 10 K).

%FIGURE 5
% Use the asterisk after figure to make the figure span 2 columns
% Use {figure} instead of {figure*} for one column
%
% /sma/ketoSci2/projects/current/h2o-dynamics/caselli5/H$_2$O_std/hnew.pro
% /sma/ketoSci2/projects/current/h2o-dynamics/caselli5/co_std/hnew.pro
%
% /sma/ketoSci2/projects/current/h2o-dynamics/caselli5/H$_2$O_over_10/hnew.pro
% /sma/ketoSci2/projects/current/h2o-dynamics/caselli5/co_over_10/hnew.pro
%
% /sma/ketoSci2/projects/current/h2o-dynamics/caselli5/H$_2$O_static/hnew.pro
% /sma/ketoSci2/projects/current/h2o-dynamics/caselli5/co_static/hnew.pro
%
% /sma/ketoSci2/projects/current/h2o-dynamics/shu/lp-h2o
% /sma/ketoSci2/projects/current/h2o-dynamics/shu/lp-co
%
% /sma/ketoSci2/projects/current/h2o-dynamics/shu/shu-h2o
% /sma/ketoSci2/projects/current/h2o-dynamics/shu/shu-co
%
%\end{figure}

\subsection{The NE-BES}
The predicted spectra for the contraction of the non-equilibrium 
sphere are shown in figure \ref{fig:spectra}. 
This model is qualitatively similar to one studied 
in \citet{FosterChev1993} except that our model 
is not isothermal and therefore not scale-free. 
However, as noted in \citet{KC10} this has a negligible 
effect on the dynamics. The predicted H$_2$O emission 
is slightly broader than in the spectrum from the QE-BES but 
negligibly different given the noise in the observed spectrum. 
The predicted CO spectrum is quite different, noticeably 
split by the higher velocities in the outer region of the NE-BES. 
Our predicted CO spectrum is essentially the same as 
the generic optically thin spectrum predicted in \citet{FosterChev1993} 
and shown in their figure 6. The only difference is that ours 
shows a slight blue-dominated asymmetry consistent with
spherical contraction. This asymmetry from self-absorption
is not expected in the optically thin approximation of the radiative transfer
used in \citet{FosterChev1993}.

\subsection{The static BES}
As a point of reference, the spectra predicted for the 
model of a static BES are shown in figure \ref{fig:spectra}.
In this case, the cold H$_2$O outside the core center 
absorbs not only the dust continuum but also all the H$_2$O
emission. The CO line is narrower than for either of the 
contracting models, and a little brighter since all
the emission is concentrated in a smaller width.

\subsection{The LP flow}
The spectra predicted for the LP flow are compared 
with the observed spectra in figure \ref{fig:spectra}.
In the LP solution there are two scaling parameters,
the sound speed and the evolutionary time that affect
the densities and velocities.
We use a
sound speed of 0.2 kms$^{-1}$, the same as in \citet{Shu1977}
and consistent with our non-isothermal models. The evolutionary
time is more arbitrary. We have chosen a time ($2.8 \times 10^{11}$ s) to best match
the observed H$_2$O and C$^{18}$O spectra. However, even the best correspondence is not good because
of the supersonic velocities in this model.
While the LP flow is a very specific model, the comparison suggests that any model of contraction
 that has a high velocity inflow in the outer part of the core is going to have trouble
matching the CO spectrum regardless of the details of the model.

\subsection{The SIS}

The spectra predicted for the inside-out collapse of the SIS  
are compared with the observed spectra in figure \ref{fig:spectra}. 
Both the H$_2$O and CO 
spectra show that the velocity profile is not quite correct.  This is not easy to fix. 
The velocity profile depends on the evolutionary time, $8\times 10^{12}$ s in our
model.
At earlier evolutionary times,  the velocity profile  
has a narrower range of velocities on the relevant core radii 
($> 100$ au). However, this also narrows the absorption feature 
in the H$_2$O spectrum and allows the redshifted emission feature
to become as bright as the blue feature. On the other hand, at later 
evolutionary times, the increased range of velocities increases the width of the CO line. 

One of the notable characteristics of the observed spectral 
line profiles in many Taurus cores is the lack of high velocity wings.  
The lack of high velocities is a critical feature that identifies the starless cores 
among the other clouds in the supersonic turbulent ISM.
The high velocities in the SIS model are problematic in this regard.

Another difficulty with the SIS model is the sensitivity of the 
width of the predicted line profiles to the evolutionary time which is
the parameter that scales the range of inward velocities. A group of
SIS cores at different evolutionary times would have individual cores
with different maximum velocities and therefore different line widths. 
This is not indicated by the observations.
Among the many spectral line observations of cores in the literature, 
we do not find a range of line widths or profiles from subsonic to supersonic as 
would be observed in SIS cores at different evolutionary times. 

In contrast, the QE-BES does not have this sensitivity to
evolutionary time. Once the central density of a quasi-equilibrium 
contracting BES reaches a certain density \hbox{($\sim 10^5 - 10^6$ cm$^{-3}$)} 
the predicted line profiles, taking account of the angular
resolution of the single-dish beams, do not change much as the 
core evolves until the velocities become supersonic in the inner core 
over a region comparable to the single dish beam size ($\sim 1000$ au).

\section{Extended inward motions in the QE-BES}

There are two other observations in the literature whose interpretation
we wish to address in the context of the QE-BES model. First, observations
of CS(2-1) show spectra asymmetrically split by self-absorption in
a line profile characteristic of inward motion \citep[figure 7 of][]{Tafalla1998}.
The splitting is seen in spectra taken at positions out to a projected 
radius of about 16000 au, assuming a distance to L1544 of 140 pc. 
This pattern is also seen in other cores in the Taurus star forming region \citep{Lee2001}.
As suggested in the references above, 
the comparison of the observed CS and N$_2$H$^+$ lines
indicates inward flow from an extended region 
defined by the CS emission onto a smaller region of
dense gas defined by the extent of N$_2$H$^+$ emission.  With the model
of the QE-BES, we can interpret these observations more precisely.

\begin{figure*}%[t]
$
\begin{array}{cc}
\includegraphics[width=5.25in,angle=90]{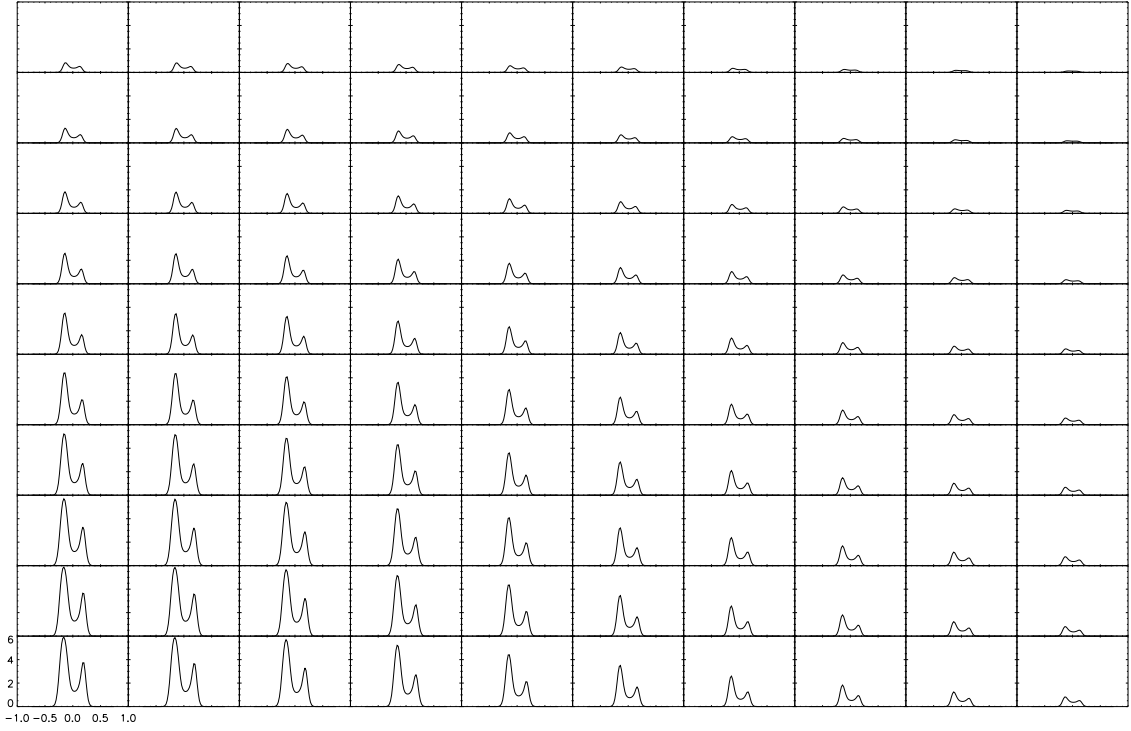}
\end{array}
$
\caption{
Map of CS(2-1) emission predicted from the model of the BES contracting in quasi-equilibrium.
The center of the model is the lower left corner. The map shows 1/4 of the spherical model.
The velocity axis ranges from -1 to +1 in the rest frame of the cloud. The brightness ranges
from 0 to 5 K.
The spacing of the spectra is  2000 au showing that the infall asymmetry is evident out to a
radius of 16000 au.
}
\label{fig:cs-map}
\end{figure*}

Figure \ref{fig:vd1b} shows that the QE-BES model for 
L1544 has inward velocities out beyond 16000 au or
100 arc seconds, again assuming a distance of 140 pc.
While these extended inflow velocities only slightly broaden the optically 
thin C$^{18}$O line (figure \ref{fig:spectra})
they cause asymmetric self-absorption in optically thick lines 
such as CS(2-1). 
The simulated CS(2-1) spectra show split 
asymmetric spectra extending almost to the boundary of the core.
Figure \ref{fig:cs-map} shows these spectra to 16000 au where the
lines become so weak that observational noise would hide the
splitting. A comparison with figure \ref{fig:vd2} shows
that these spectra are generated within the $1/r^2$ region of the QE-BES. 

% FIGURE 10

The difference between the spatial extents of the CS and N$_2$H$^+$ emission and
the difference in their spectral line profiles are easily understood as
arising from their respective abundances and variations in abundance with density.
The CS molecule is about 100 times as abundant as 
N$_2$H$^+$ \citep{vanDishoeck1998,KC10}. Since both molecules have
similar Einstein A coefficients and similar critical densities for
collisional de-excitation, the greater abundance of CS makes this molecule
brighter than N$_2$H$^+$ in lower density gas.

Based on our earlier research \citep{KC10}, we assume that all the N$_2$H$^+$ 
remains in the gas phase across the core. For our purposes, this is consistent with \citet{Caselli2002} 
who found that N$_2$H$^+$ was
depleted by a factor of 3 in the center of L1544. This is much smaller
than the depletion of CO which is almost complete at the highest densities in L1544.

Our simple chemical models do not include sulfur species. 
In the cold cores, the dominant processes controlling the abundance are
photo-dissociation and freeze-out which we assume to be similar for CS and CO. 
We know that there are differences in the gas phase chemistry of CS and CO particularly
at high densities, but  because the gas phase
abundance of both molecules is decreasing rapidly by freeze-out at these same densities, 
the spectral lines of both molecules are dominated by the emission from the outer core
where the abundances are similar.
However, the maximum 
brightness in our CS line is just under 5 K, a little brighter than the peak brightness
in the observed line, just under 4 K.

If we assume that
most of the CO in L1544 is in C$^{16}$O, and we adopt the C$^{16}$O/C$^{18}$O ratio of 560
suggested by \citet{WilsonRood1994}, this fixes the total carbon abundance relative
to H$_2$ at $1.3\times 10^{-4}$ and the maximum CS abundance (neither depleted 
nor photo-dissociated) at $3 \times 10^{-8}$ relative to H$_2$. However, because
the CS line is optically thick, the intensity is not very sensitive to the abundance.

\begin{figure}%[t]
$
\begin{array}{cc}
\includegraphics[width=3.25in]{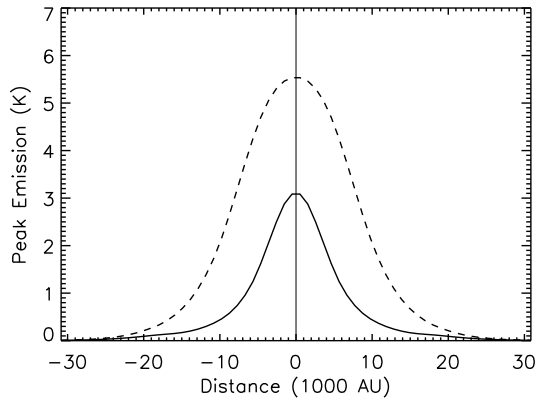}
\end{array}
$
\caption{
Comparison of the modeled spatial extent of the CS(2-1) (dashed lines) and 
N$_2$H$^+$(1-0) (solid line) emission across the L1544 core. The difference is
mostly due to their different abundances. CS is about 100 times more abundant than
N$_2$H$^+$.
}
\label{fig:comparison}
\end{figure}

Figure \ref{fig:comparison} shows the spatial variation of the peak 
brightness of CS(2-1) and N$_2$H$^+$(1-0) in the QE-BES model
as calculated by our radiative transfer simulation.
The simulation shows 
that in the QE-BES model, the extent of the CS
emission is twice that of the N$_2$H$^+$. The spatial half width at half maximum (HWHM)
of the simulated peak emission of the CS(2-1) is 8200 au or 0.04 pc. This
may be compared to the observed HWHM of 10300 au or 0.05 pc  \citep{Lee2001}.
The width in the CS simulation includes convolution with a 48" FWHM beam to match the 
beam of the observations. The spatial HWHM of the modeled N$_2$H$^+$ emission is
5100 au or 0.025 pc. This may be compared to an observed HWHM for the N$_2$H$^+$ emission of
5000 au or 0.025 pc  \citep{Lee2001} and 6000 au or 0.03 pc \citep{Caselli2002}.
The difference in the 
extent of the CS emission between the model and the observations may be
due to the approximation of spherical symmetry in the model whereas L1544
is observed to have an axial ratio of 2:1. The observationally determined radius includes
this spatial asymmetry since it is measured
as $\sqrt{A/\pi}$ where $A$ is the area of the cloud within
the HWHM contour. Our
approximate treatment of the CS abundance may also contribute to this
discrepancy.

Since N$_2$H$^+$ has a nearly constant abundance (we assume $3\times 10^{-10}$)
inside the photo-dissociation boundary, its line brightness scales with
the column density. However, owing to its low abundance, the N$_2$H$^+$ line is 
optically thin except in the very center of the core where the 
hyperfine lines are seen to be just barely split by self-absorption\citep[figure 9 of ][]{KC10}. 
The smaller splitting does not indicate lower velocities in the inner core than in
the outer, just lower optical depth of the molecular tracer. In general the magnitude
of the inward velocities cannot be inferred only from the width of the splitting due to
self-absorption. The observed separation of the split peaks is due to a combination of the velocity and the
optical depth so the strength of the self-absorption must be known as well.

It is correct to infer that the bright N$_2$H$^+$ emission arises from the densest 
part of the core and that there is contraction from the outer core  onto this dense
region. However, the observations of the H$_2$O line indicate even higher
contraction velocities in the dense inner core.
Furthermore, the CS and N$_2$H$^+$ spectra do not indicate inward velocities
extended beyond the core. The inward velocities responsible for splitting the CS
spectra are entirely within the core.

Another simple line of reasoning leads to a similar conclusion.
We know that L1544 is bounded by a photo-dissociation region that is
necessarily created by the external UV radiation that is required to produce
the outwardly increasing gas temperature \citep{Evans2001,Crapsi2007} and the
gas phase abundance of H$_2$O resulting from photo-desorption \citep{KRC14}. 
Setting aside the complex radiative transfer arguments, we know that 
the molecular line emission must 
come from the molecular side, the inside, of the photo-dissociation boundary. 

Alternatively, the outer boundary can also be defined observationally as a 
transition to coherence \citep{Goodman1998,Pineda2010}.  
The observed width and splitting of the CS line indicates subsonic velocities. 
This places the CS emission
within (on the subsonic side) of the transition to coherence. 
These considerations again place the extended inward motions within 
and not outside the QE-BES.

As correctly pointed out in \citet{Tafalla1998}, inward velocities at the 
radii observed in L1544 are incompatible with the inside-out collapse 
of an SIS.  In the SIS model, non-zero velocities at
radii of 16000 au require an evolutionary time $\simge 0.4$ Myr.
This is  the time required  for the expansion wave initiating the 
infall to reach 15000 au. During this time,
enough mass would have accreted into the center to form an easily
detectable protostar \citep{Tafalla1998}. The QE-BES model does not have this inconsistency.
In contrast to the SIS model, the central density in the  
QE-BES is only $10^7$ cm$^{-3}$ at the time 
that the inward flow extends beyond 30000 au (figure \ref{fig:vd1a}).

There is an interesting difference between the accretion rates of the SIS and QE-BES
models.
Whereas the accretion rate in the SIS is constant in time, the accretion
rate in the QE-BES model is initially zero and increases with time. However,
the characteristic time scale for collapse is essentially the same
in both models and is the
free-fall or sound-crossing time. This comparison implies that
in the QE-BES contraction, the formation and
accretion onto a protostar occurs later than in the SIS, still within 
the collapse time of the entire core, but later within 
that time period.  

% FIGURE 11
% /sma/ketoSci2/projects/current/h2o-dynamics/caselli5/comparison-cs-n2h

\section{Speculation on the origin of the QE-BES}

Our study seeks to better understand how the clouds that we observationally identify as starless cores
evolve toward star formation. The formation of the starless cores within the ISM is a topic for further
research. However, we can speculate on the more general question of whether an unstable 
core such as the QE-BES is a suitable initial state for an evolutionary model. 
How is this unstable state created? Is this model subject to the same criticism as directed toward the SIS
that this initial state is unrealizable because of its instability \citep{Whitworth1996}? 
We imagine that an unstable BES may be 
created in a two
step process. The first step is the creation of a stable BES out of the larger scale turbulent ISM and 
the second step is the evolution of the stable BES to instability.  The second step is better understood
and can be explained rather easily if we assume that stable cores can be created. We begin with the
second step first.

\subsection{The approach to instability}

One process that leads a stable BES to instability, the dissipation of internal turbulent 
kinetic energy, was recognized immediately in the early observations that first identified the
star-forming potential of the starless cores.  
"Observed distributions of dense core properties in Taurus, Ophiucus, and other dark cloud 
regions appear consistent with dense core evolution toward star formation via 
dissipation of turbulence \citep{Myers1983}".  
The internal subsonic turbulent kinetic energy is critical to 
the dynamical stability of the cores because thermal energy provides only a fraction of the
internal energy needed to support a core against gravity and external pressure.
For example, \citet{MyersBenson1983} suggested that the 
line widths of starless cores, equivalent to a 16 K Gaussian, show sufficient
internal energy for virial equilibrium, but that the 10 K gas could account for only 60\% of
the kinetic energy indicated by the line width. 
The early observations also noted that star formation activity was associated with cores with
less subsonic turbulence. From observations of 27 cores from \citet{MyersBenson1983} and 
16 cores from \citet{LKM1982}, \citet{Myers1983} concluded
"In complexes 
which are vigorously forming low-mass stars, dense cores are more prevalent, 
smaller, denser, and have narrower lines than in regions with less star formation. "
These results were confirmed in a much larger survey of 179 cores by \citet{Tachihara2002}
who noted smaller molecular line widths in star-forming cores than in starless cores. They also
concluded that decay of the internal subsonic turbulence is necessary for star formation.
More recently, a survey of cores in the Pipe nebula also shows 
many cores requiring turbulent energy in addition
to thermal energy for equilibrium \citep{Lada2008}.

On the theoretical side, recent research \citep{Keto2006,Broderick2007} 
identifies the complex spatial patterns 
of red and blue asymmetric self-absorbed spectral line 
profiles observed in some cores \citep{Lada2003,Redman2006,Aguti2007} 
as the long-wavelength modes of this turbulence or sonic oscillations in BES.
On the basis of numerical simulations \citet{Broderick2007} 
concluded that long-wavelength oscillations,
except for the (0,0,0) breathing mode, can indeed stabilize a BES. Analysis of wave 
transmission and reflection as a three-wave problem \citep{Broderick2008} and non-linear
mode-mode coupling by numerical
simulations \citep{Broderick2010} indicates damping times for oscillations in BES
of several $10^5$ to just over $10^6$ yr depending on the density contrast between the 
sphere and its surroundings.

The transition from a stable BES to an unstable QE-BES can be summarized as follows.
For a given internal energy, including thermal and subsonic turbulence, expressible as an
effective sound speed, and a given external pressure
there is a maximum stable mass for a BES. As the turbulent energy dissipates, the critical
mass lessens until it becomes equal to the actual mass. At this point the BES is at critical 
equilibrium and ready to begin its contraction as an unstable QE-BES 
with any subsequent loss of internal energy.

\subsection{The formation of stable cores}

On scales larger than the cores, the ISM is supersonically turbulent, yet the character
of that turbulence is described differently in the literature. Numerical hydrodynamic 
simulations of the turbulent ISM generally suppose that the turbulence has an external 
driver that has produced the initial turbulent velocities and densities
\citep{Klessen2005,Offner2008}. More conceptual models for the supersonic turbulence
emphasize the dominance of internal gravitational rather than hydrodynamic forces in the 
supersonic turbulence of the ISM \citep{Larson1981,FBK2008}.

The numerical hydrodynamic simulations of the turbulent ISM have not yet produced
stable cores. These results are interpreted differently. \citet{Klessen2005} and \citet{BP2003}
suggest that the transient density fluctuations in their simulations of the turbulent ISM 
resemble BES and the interpretation of actual observations as indicating BES in the ISM
is mistaken. 
However, \citet{Offner2008} interpret their own simulations otherwise:
"The physical origin of the poor agreement between the simulations and observations appears
to be that the simulated protostellar second-moment distributions ...  do not 
have sufficiently narrow peaks. The protostellar cores in the simulations are at the centers of regions of
supersonic infall, which contradicts the observations that show at most transonic contraction."

A discussion of the numerical hydrodynamical modeling of turbulence is beyond the scope of this
paper other than to note the differences in opinion among numerical modelers. 
The very strong conclusion that  "Reaching hydrostatic equilibrium in a turbulent molecular 
cloud environment is extremely difficult and requires strongly idealized conditions that are not met 
in the interstellar gas \citep{Klessen2005}" is not shared by all researchers of the turbulent ISM.
For example, the more conceptual model of the ISM turbulence of \citet{FBK2008} suggests how the 
stable starless cores might be created.

\citet{FBK2008} imagine the ISM as a turbulent cascade of mass and energy from larger to
smaller scales. The transfer takes place through fragmentation of larger scale clouds which
is motivated by the decay of the turbulence. In this picture the gravitational and
kinetic energies of the clouds in the ISM are in virial equilibrium with an external pressure.  The concept of
a maximum mass for stable equilibrium applies to these clouds in virial equilibrium the same way
as it does to the BES in hydrostatic equilibrium. As with the BES, the larger scale clouds continuously
lose their internal kinetic energy through dissipation of their turbulence. Unlike the BES,
these larger clouds whose dynamics are dominated by the inertial forces of the supersonic turbulence
can fragment into smaller clouds whose masses are below their maximum stable masses, at least
for the moment until they too lose energy through turbulent dissipation. This process continues
to smaller and smaller scales until the clouds are small enough to be supported by subsonic 
turbulence and thermal pressure. This is the sonic scale and the scale of the starless cores.

This process imagines that cores with subsonic turbulence are created in a range of masses above 
and below the maximum stable mass. Those cores with higher masses immediately begin 
contraction to star formation. These are short-lived and rarely observed. We are left with cores whose
masses can be supported by a combination of thermal and subsonic turbulent energy. These
are the cores we see in the Taurus star forming region  
evolving to star formation through the decay of their subsonic turbulence.
Other cores whose masses are small enough to be supported by thermal energy alone may
never form stars. Many of the cores in the Pipe nebula are in this category \citep{Lada2008}.

\section{Discussion}

The conclusion of our research on L1544 is that this core is contracting in a way that is very
specifically matched by the contraction of a QE-BES. The comparison with the NE-BES indicates
that even a 10\% deviation from (unstable) equilibrium will create contraction velocities
incompatible with the observations. Other modes of contraction with supersonic velocities
either at large or small scales, for example, but not limited to, the LP flow or the SIS inside-out collapse,
are also precluded by the data. In this sense, the LP and SIS models serve as examples to stand for 
many other types of models with supersonic flows. For example, the LP model with its supersonic 
velocities at the cloud boundary
may be representative of other models that form the starless cores by compression of larger-scale 
flows or models of cores that accrete gas from larger scales. The NE-BES model may be 
representative of many types of models characterized by free-fall collapse. The SIS model may be 
representative of many types of models that lack pressure support at the center, for example 
accretion onto  a gravitational point source or a pressure-free sink cell.
In this sense, although the several models considered in this study are quite specific, they suggest
against many other types of models for the starless cores particularly those with supersonic velocities 
either at small or large scales.

Stressing the conditionals here, if it is hard to imagine that L1544 can 
so closely resemble a quasi-equlibrium hydrostatic cloud without actually being one, and if
it is hard to imagine conditions in the turbulent ISM that can create hydrostatic cores, 
then our conclusion of the quasi-hydrostatic state of L1544 goes head-to-head against 
the strong conclusion from some of the numerical hydrodynamical
simulations that hydrostatic clouds cannot be created in the turbulent ISM. 

However, we have so far studied only this one core at this level of detail, and the observation of the 
water emission line in L1544 is unique among starless cores. While L1544 shows no indication that it is unique 
among the cores in Taurus, definitive conclusions obviously should not be drawn from this one example. 
Further observations of other cores are needed. Because the Herschel Space Observatory is no
longer operating, we will have to find other tracers that isolate the conditions in the core centers. 
The deuterated species of some molecules are possibilities \citep{VDT2005,Caselli2003}.

\bigskip
P. Caselli acknowledges the financial support of the European Research Council (ERC; project PALs 320620)

\bibliography{ms1}

\end{document}